\documentclass[
rsi,
 amsmath,amssymb,
 reprint,%
]{revtex4-1}

\usepackage{graphicx}
\usepackage{dcolumn}
\usepackage{bm}
\usepackage{float}
\usepackage[utf8]{inputenc}
\usepackage[T1]{fontenc}
\usepackage{mathptmx}
\usepackage{etoolbox}
\usepackage{mathtools}
\usepackage[svgnames,table]{xcolor}
\usepackage{subcaption}
\usepackage{multirow} 
\usepackage{array}

\newcolumntype{L}[1]{>{\raggedright\let\newline\\\arraybackslash\hspace{0pt}}m{#1}}
\newcolumntype{C}[1]{>{\centering\let\newline\\\arraybackslash\hspace{0pt}}m{#1}}
\newcolumntype{R}[1]{>{\raggedleft\let\newline\\\arraybackslash\hspace{0pt}}m{#1}}

\usepackage{hyperref}
\hypersetup{
        colorlinks=true,
        citecolor=SteelBlue,
        filecolor=LimeGreen,
        linkcolor=SlateBlue,
        urlcolor=MediumPurple
}

\makeatletter
\def\@email#1#2{%
 \endgroup
 \patchcmd{\titleblock@produce}
  {\frontmatter@RRAPformat}
  {\frontmatter@RRAPformat{\produce@RRAP{*#1\href{mailto:#2}{#2}}}\frontmatter@RRAPformat}
  {}{}
}%
\makeatother
\begin{document}

\preprint{AIP/123-QED}

\title[ Demonstration of the minimal coupling of horizontal acceleration to rotation in a torsion balance suspended from three-wires]{Demonstration of the minimal coupling of horizontal accelerations to rotations in a torsion balance suspended from three wires}
\author{Amit Singh Ubhi}
\affiliation{School of Physics and Astronomy, and Institute for Gravitational Wave Astronomy, University of Birmingham, Birmingham B15 2TT, United Kingdom}

\author{Clive C. Speake}
\affiliation{School of Physics and Astronomy, and Institute for Gravitational Wave Astronomy, University of Birmingham, Birmingham B15 2TT, United Kingdom}

\author{Emilia Chick}
\affiliation{School of Physics and Astronomy, and Institute for Gravitational Wave Astronomy, University of Birmingham, Birmingham B15 2TT, United Kingdom}

\author{Conner Gettings}
\affiliation{Center for Experimental Nuclear Physics and Astrophysics, University of Washington, Seattle, WA 98195, United States of America}
%

\date{\today}

\begin{abstract}
The Cavendish torsion balance is the instrument of choice for measuring weak forces, such as gravity. Although torsion balances  have extremely high sensitivity for measuring forces over ranges of a few cm and more, their dynamics make it difficult to extend this range to much less than fractions of mm.   In particular forces such as the Casimir force are usually studied using atomic force microscopes. We present results of our studies of a simple torsion balance with a 3-wire suspension. This device should be able to maintain parallelism between flat plates of areas of a few $\sim\mathrm{cm^2}$ at  separations of much less of 10's of $\mathrm{\mu m}$. In this paper we describe our experimental investigation into the coupling of ground tilt to the torsional rotation of the novel device. We show that, like the Cavendish torsion balance, the 3-wire torsion balance is highly insensitive to tilts. We also demonstrate that this tilt sensitivity is itself insensitive to shifts in the centre of mass position of the suspended mass. We discuss simple models of the 3-wire torsion balance that show that it is only the static wire lengths that determine the coupling of tilts and horizontal accelerations. We also discuss designs of torsion balances where sensitivity and noise rejection to tilts are optimised.
\end{abstract}

\maketitle

\section{Introduction}\label{sec:Introduction}

In our laboratory we have been developing devices that are intended to allow weak forces to be detected between objects in close proximity. Such detectors are forms of torsion balances and could be used to measure, for example, Casimir forces or search for deviations of the inverse square law of gravitation. We have reviewed the motivations of this work in our previous papers \cite{SpeakeCollins,GettingsSpeake}. In brief, Cavendish recognised that Michel's torsion balance made it possible to accurately measure forces acting on a test mass that are significantly smaller than their weight. This is because the torsion balance detects torques acting about the local gravitational vertical. These torques are due to the forces acting on the suspended test masses that are perpendicular to local gravity and the weights of the suspended test objects. Torsion balances respond to time varying tilts and horizontal accelerations of the lab frame, produced by ground motion, with simple pendulum oscillations.  However these oscillations couple weakly to the torsional oscillation mode, which contains the signal of interest, because the centre of mass of the suspended object lies essentially on the axis of rotation. This is true provided the torsion bob has at least three-fold symmetry \cite{SpeakeGillies}. It is thought that the sensitivity of Cavendish-type torsion balances to static ground tilt is due to departures from axial symmetry of the fibre at its upper attachment point. 
Hoyle, for example, \cite{HoyleThesis} reports that a tungsten fibre of 20 $\mu$m diameter and 80 cm length twisted by an angle corresponding to   1.6 $\%$ of the tilt applied at the suspension point. This fibre was attached to pre-hanger that used another tungsten fibre of  20 $\mu$m diameter and 3 cm length as a means of reducing the tilt/twist coupling. This mechanism could potentially give rise to tilt to torsional coupling in the device discussed in this paper.

In Ref.\,\cite{SpeakeCollins} we discuss the general principles of the construction of devices that do not rotate due to horizontal accelerations or tilts but also have dynamics that can be more easily controlled. This is achieved by matching the centre of mass of the suspended object with the centre of stiffness, or buoyancy of the suspension. This idea was originally conceived in the context of a superconducting torsion balance \cite{SpeakeCollins} and was then demonstrated convincingly with an air suspension that was controlled by magnetic actuators \cite{GettingsSpeake}. In this paper we describe our recent progress in this endeavour and present a simple system that achieves this matched condition passively. The device described here could form the basis of a new instrument that would allow forces with short ranges to be measured precisely. 

It is important to note that, as further discussed in Sec.\,\ref {sec:PhysicalPrinciples}, in order to benefit from the insensitivity of the 3-wire torsion balance to tilt, we require measurement schemes which are insensitive to the translation of the suspended test object that accompany any horizontal or tilt ground motion. Such schemes are used on Cavendish-type balances and include autocollimators \cite{ShawAdelbergerTorsionBalance,LeeAdelbergerISL}, or custom-built rotation sensors. We have constructed an interferometric device for measurement of angle that we intend to use on the final device \cite{ILIAD}. In this current work we use a simple combination of an optical lever and lens to achieve a readout with nominally zero sensitivity to the translation of the test object. We describe this in Sec.\,\ref{sec:experimental_design}.

\section{Physical Principles}\label{sec:PhysicalPrinciples}

\begin{figure}[!htbp]
    \centering
    \includegraphics[width = \columnwidth]{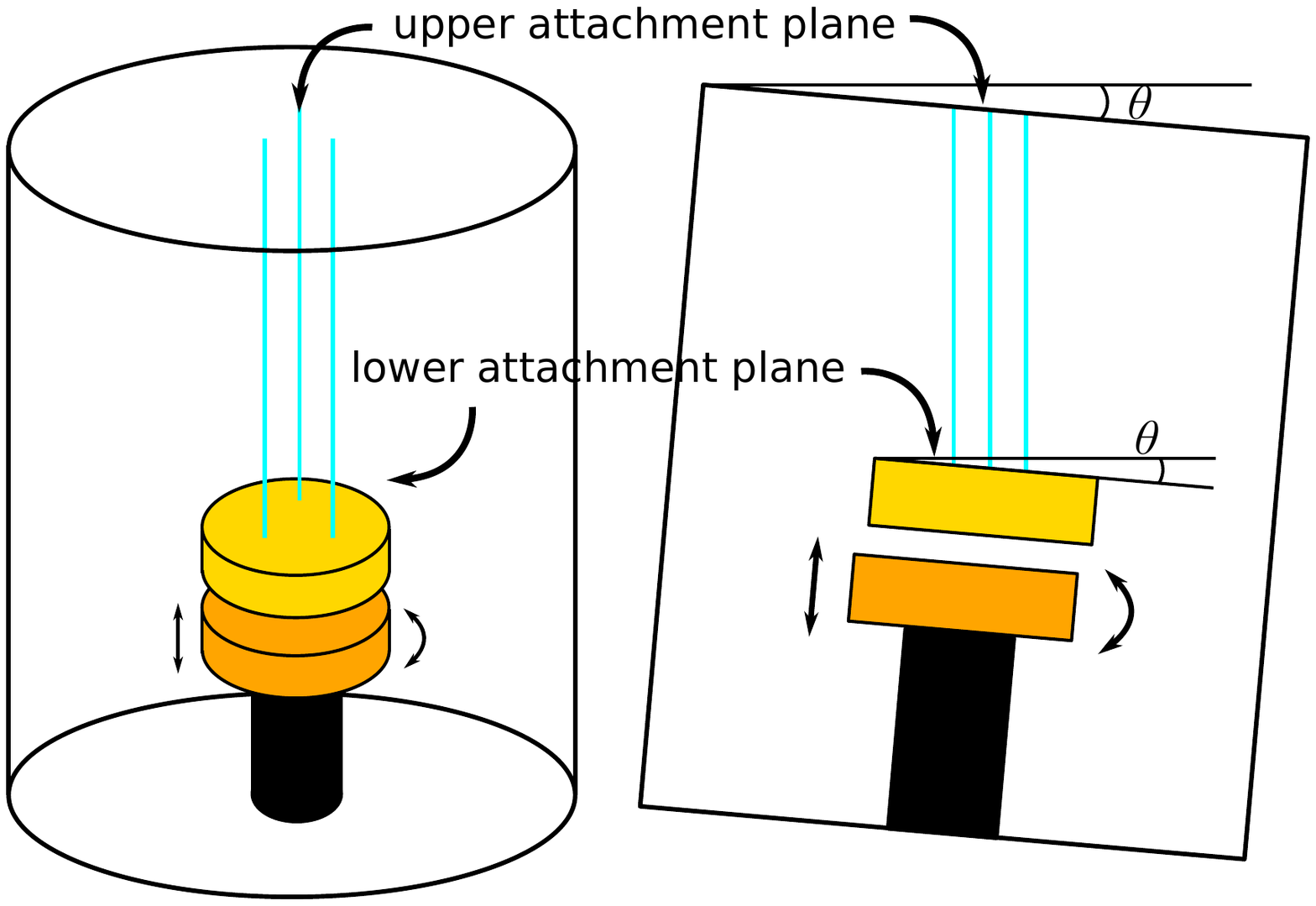}
    \caption{\textbf{Left:} Isometric diagram of a mass suspended via three wires from a frame with a secondary interacting mass below. \textbf{Right:} When the frame is tilted, the three wires constrain the suspended mass such that it remains at the same angle relative to the frame, resulting in the two interacting masses remaining in the same angular orientation with respect to each other.}
    \label{fig:3wires}
\end{figure}

We describe the detailed physical and mechanical characteristics of the device in Sec.\,\ref{sec:experimental_design}, here we discuss the general principles of its design.
A schematic of the new device is shown in Fig.\,\ref{fig:3wires}. Three wires are attached to an upper plate and these support a lower plate. The orientation of the lower plate will be determined by the positions and lengths of the wires provided that the upper and lower points of attachment are not co-linear. We also assume here that the pattern of the upper points of attachment in the upper plate match those of the lower plate, such that the wires are parallel to each other.

To understand the basic principle of the new torsion balance it is useful to think of it as a system of three individual pendulums of equal length that are suspended from the top-plate using ideal pivots. The three pendulums support, respectively, a point mass (or bob) of value $m_1$, $m_2$, and $m_3$, which are connected by light, rigid and inextensible rods. The mass values produce tensions in the pendulums corresponding to the loads that the pendulums would support for a lower plate of arbitrary mass distribution. It is obvious that a tilt of the top plate does not result in relative horizontal displacements of the pendulum bobs as each pendulum will remain parallel to the local gravity field. Thus it can be deduced that, provided that the pendulums are the same length, the lower plate will not rotate, irrespective of its mass distribution. Throughout this paper we will limit our discussion to the low-frequency dynamics (or DC response) of the system, hence the moments of inertia of the device, for example, do not need to be considered.

If the frame from which the system is suspended is tilted about any horizontal axis, an observer would note that there is translational motion of the suspended mass (tilt-to-translational coupling discussed in Ref.\,\cite{Matichard2015, Ubhi6D, UbhiPlatform, Prokhorov6D}) but no rotation of the suspended mass about a vertical axis occurs. This device therefore provides a degree of passive isolation to seismically induced motion in this rotational degrees of freedom, as is the case of the Cavendish torsion balance. The novelty of this system is that the torsion balance mass (the suspended plate in our discussion) remains parallel to the top plate from which it is suspended. As noted above, this allows the possibility of measuring forces between parallel surfaces where the separation between the objects can be minimised. 

Another way of approaching this problem is to consider, for simplicity, two vertical wires that support a horizontal rod with a non-uniform mass distribution. This could be thought of as a trapeze with a gymnast who is not sitting at its centre. The tensions in the wires can be found, in the usual way, by balancing the forces and torques due to these tensions with their forces and torques due to the weight supported. Then consider the case where a horizontal acceleration is applied at low frequency to the attachment points of the wires perpendicular to the trapeze. The resulting acceleration on the trapeze acts at its centre of mass in the opposite direction to the applied acceleration. Carrying out the same balancing procedure of the forces and moments, but now in the horizontal plane, shows that the displacement of the rod is purely translation, with no rotation. This is guaranteed because the tensions in the wires give the horizontal components of force that are proportional to the horizontal displacement of the rod relative to their points of attachment. The magnitudes of these tensions themselves depend on the location of the centre of mass of the supported object, and this ensures that the displacements of the wires at each end of the trapeze are equal. If the gymnast sits closer to one wire, this wire has a larger tension and produces a larger stiffness against forces in the horizontal plane than does the other wire. This simple system passively decouples tilt from rotation. It is an example of a system that satisfies the general requirements that we have previously described in the context of superconducting suspensions \cite{SpeakeCollins}.

The mathematical solution to this problem is discussed here for a generalised $n$-wire suspension. The system is solved via three simultaneous equations, modelling the static equilibrium positions of each displaced position of the pendulum bobs in the horizontal plane. The  positions can be written as $x$, $y$, and $\phi$, which are translations of the lower plate in $\hat{x}$, $\hat{y}$, and rotation about the $\hat{z}$ directions respectively, in a cartesian coordinate system. For convenience we can consider the centre of this coordinate system to have its origin at the geometrical centre of undisplaced lower plate.

The pendulum bobs are connected to the centre of the lower plate by light rods of nominal length $R_{i}$ and the cartesian coordinate of the $i$th bob at its initial position is $\underline{r_{i}}^{\mathrm{init}}=R_{i}\cos\theta_{i}\hat{x}+R_{i}\sin\theta_{i}\hat{y}$. We can define the new position of each bob as $\underline{r_{i}}^{\mathrm{disp}}=(R_{i}\cos(\theta_{i} + \phi) + x)\hat{x}+(R_{i}\sin(\theta_{i} + \phi) + y )\hat{y}$. The difference in positions between the displaced and initial positions of each bob is $\underline{dr_i} = \underline{r_{i}}^{\mathrm{disp}} - \underline{r_{i}}^{\mathrm{init}}$. A tilt of the system is equivalent to an acceleration applied at the centre of mass, or in the case of modelling the suspended mass as individual bobs, an equal acceleration applied to each bob. The tilt can then be modelled by forces acting on each bob,

\begin{equation}
    \underline{F}=\sum_{i}\underline{F_{i}} = \sum_i m_i \underline{a} = \sum_{i}m_{i}\left(a_{x}\hat{x}+a_{y}\hat{y}\right).
\end{equation}

The three equations describing the displaced equilibrium position due to the applied accelerations are, the static force balance in the $\hat{x}$, and $\hat{y}$ directions,

\begin{equation}
    \sum_i \left[ -\frac{m_i g}{L_i}\underline{dr_i} + m_i\underline{a}  \right] = \underline{0}
\end{equation}

where $m_i g/L_i$ is the horizontal translational stiffness of the $i$th bob, and $L_i$ is the respective wire length. The condition on the torques about the $\hat{z}$ axis becomes,

\begin{equation}
    \sum_i \left[ \underline{r_{i}^{\mathrm{disp}}} \times \underline{F_i} + \underline{r_{i}^{\mathrm{disp}}} \times -\frac{m_i g}{L_i}\underline{dr_i} \right] = \underline{0}
\end{equation}

Solving these static equations of forces and torques, it is found that with equal wire lengths, the torsional angle $\phi$ is always zero regardless of any other asymmetries in the system such as tensions, wire positions and differences between wire positions on the top and bottom plates. We have modelled deviations from the ideal scenario as variations in the wire lengths. Even with imperfect symmetry, the torsion angle is linearly dependent on the differences in the wire lengths. Differences in the lengths of the wires allows asymmetries in the geometry and tensions to couple linearly to the rotation as higher order effects (i.e. effects that depend on the product of a wire length difference and another parameter describing an asymmetry). Eqn.\,\ref{eq:torsionangle} in Sec.\,\ref{sec:procedure} gives the relationship between rotation and tilt for differences in the wire lengths. Note this calculation does not include the axial asymmetry of the individual wires which is the general mechanism of tilt coupling for Cavendish-style torsion balances as mentioned in Sec.\,\ref{sec:Introduction}.

\section{\label{sec:experimental_design}Experimental Design}

\begin{figure}[!htp]
    \centering
    \includegraphics[width = \columnwidth]{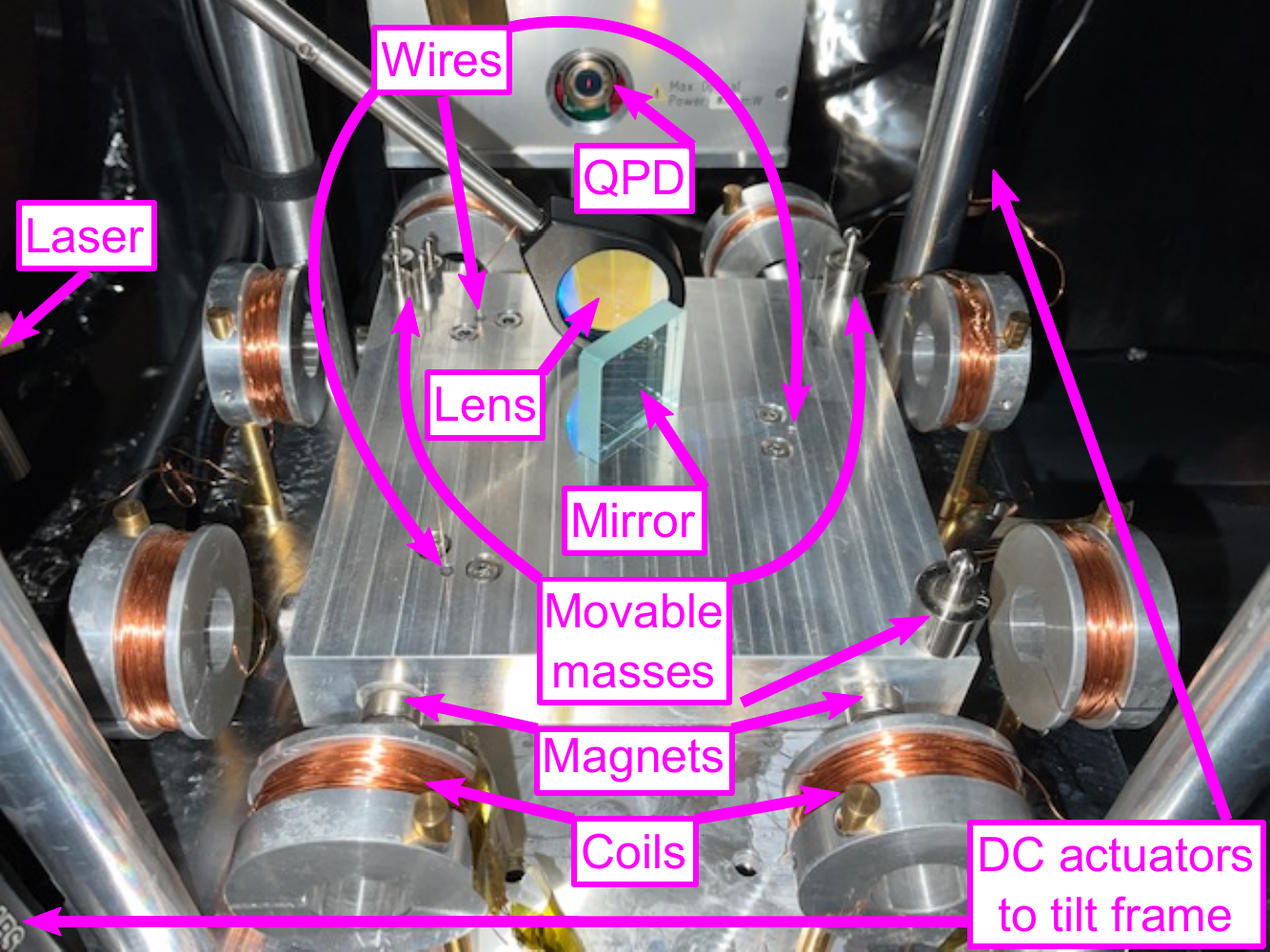}
    \caption{Labelled photo of the experimental apparatus.}
    \label{fig:Photo}
\end{figure}

\begin{table*}[t]
\caption{Parameters for the three wire torsion balance.}

\begin{ruledtabular}
\begin{tabular}{C{2cm}|C{9cm}|C{6cm}}
Parameters & Description & Value\\
\hline
$M_{\rm tb}$ & Mass of torsion balance & $339 \pm 5$\,g \\
$M_{\rm add}$ & Mass of additional weight & $10 \pm 0.01$ \,g \\
$R_1$ & Radial distance to wire 1 from centre of plate & $50.9 \pm 0.5$\,mm \\
$R_2$ & Radial distance to wire 2 from centre of plate & $50.9 \pm 0.5$\,mm \\
$R_3$ & Radial distance to wire 3 from centre of plate & $33.3 \pm 0.5$\,mm \\
$R$ & Radial distance to wires from geometric centre & $44.5 \pm 0.5$\,mm \\
$w$ & Mass width (side length) & $115 \pm 0.5$\,mm \\
$d$ & Offset between geometric centres = $R/4$ & $11.1 \pm 0.5$\,mm \\
$L_{\rm{w}}$ & Wire length & $200 \pm 1$\,mm \\
$r_{\rm{w}}$ & Wire radius & $40 \pm 4 \,\mu\rm{m}$\\
$E$ & Wire Youngs modulus & 125\,GPa \\
$l_f$ & Lens focal length & $100 \pm 2$\,mm \\
$k$ & Torsional stiffness & $0.0308 \pm 0.0005$\,Nm/rad \\
$f_0$ & Torsional resonant frequency & $0.879 \pm 0.001$\,Hz \\

\end{tabular}
\end{ruledtabular}
\label{tab:params}
\end{table*}

\begin{figure*}[t]
    \centering
    \includegraphics[width=0.95\textwidth]{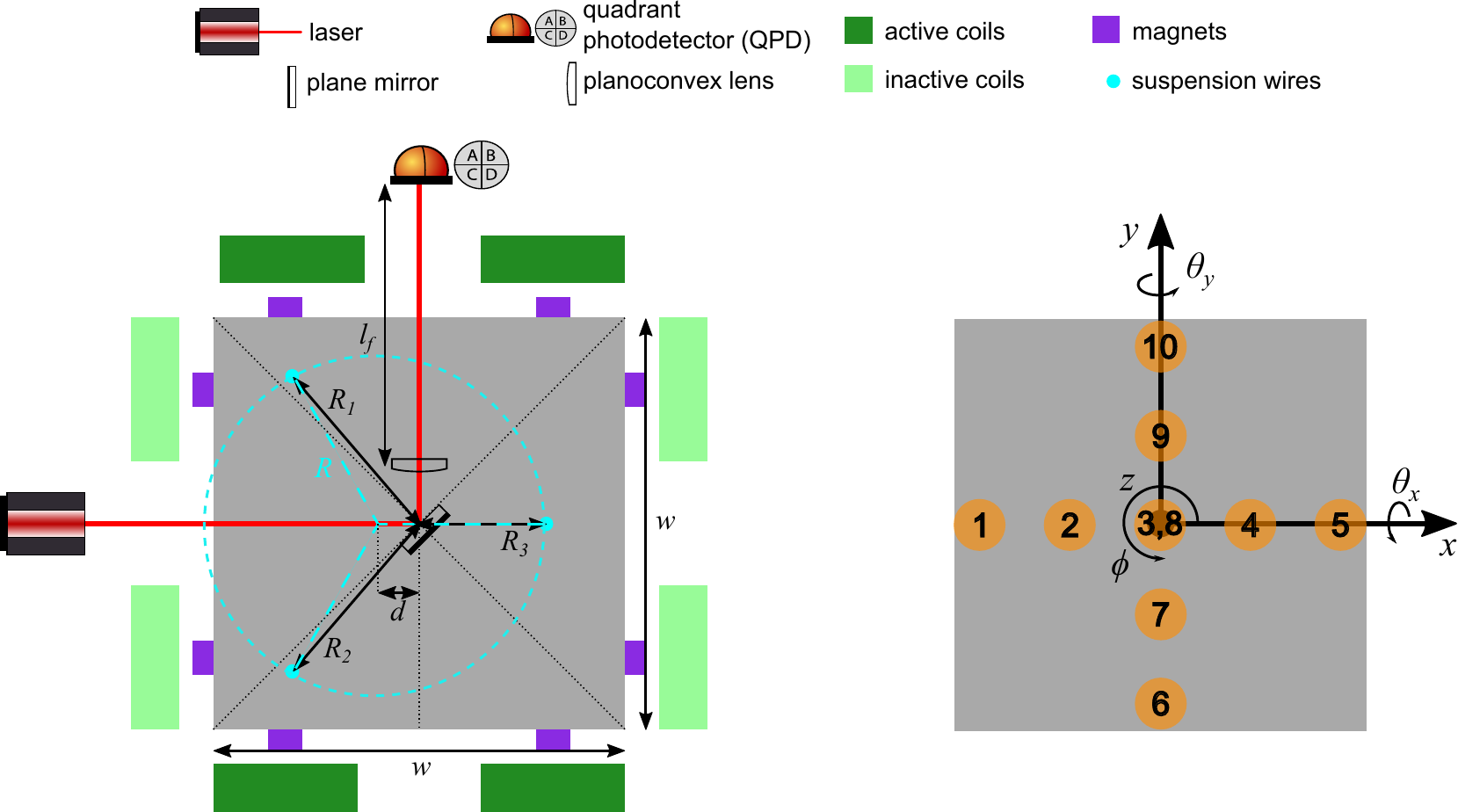}
    \caption{A schematic of experiment without the frame. \textbf{Left:} Arrangement of optical layout, actuators for feedback control, and wire positions. Key dimensions are labelled and their values shown in Tab.\,\ref{tab:params}. The geometric centre of the wires is displaced by a distance $d$ from the centre of the suspended plate (the approximate centre of mass). The distances $R_1,R_2,$ and $R_3$ are measured from the centre of the plate. The distance $R$ is the radial distance of the wires from the centre of the dotted blue circle. The QPD is placed at a distance equal to the focal length from the lens. \textbf{Right:} The coordinate origin is placed at the centre of the suspended plate in its equilibrium position, with right handed rotations. The equivalent positions of the mass are shown which is achieved by using multiple movable masses as indicated in Fig.\,\ref{fig:Photo}.}
    \label{fig:schematic}
\end{figure*}

We have replaced the air-bearing described in Ref.\,\cite{GettingsSpeake} with a suspension comprising three vertical wires shown in Fig.\,\ref{fig:Photo}. The new apparatus was located in a  multi-purpose vacuum chamber. The entire system is constructed relative to a frame, with the wires attached to a flat top-plate at the top of the system. The wires were made from copper beryllium and clamped between molybdenum rollers to minimise the contact area of the connection, with the aim of reducing energy loss in the suspension \cite{Cagnoli1996}.

The wire lengths were set by using a spacer between the upper and lower plates during its construction to achieve parallelism between them, and therefore, equal lengths of the three wires. This method was not robust as the tensions applied during wire clamping were not well determined. The wires were orientated in an equilateral triangle inscribed in a circle of radius $R$, separated by $120^\circ $. The geometric centre of this triangle was deliberately offset from the geometric centre of the mass that it suspends by a distance $d$ in the x-axis. This introduced a large difference in the tension of one of the wires. The parameters of the suspension and an illustration of the geometry are shown in Tab.\,\ref{tab:params} and Fig.\,\ref{fig:schematic}, respectively. The tension in wire 3 was twice that of wires 1 and 2. This would correspond to a system where the lower plate were replaced with light rigid rods coupling the lower ends of the wires, as discussed in Sec\,\ref{sec:PhysicalPrinciples} with masses $m_1 = m_2 = m$, and $m_3 = 2m$, where the mass of the suspended plate is $4m$. Using the equations given in Sec.\,\ref{sec:PhysicalPrinciples}, we see that horizontal accelerations/tilts do not induce rotations of the mass regardless of the variations in $m_1, m_2$ and $m_3$ as long as the wire lengths remain equal. 

The total torsional stiffness, $k$, of the device is determined by summing the individual horizontal translational stiffness contributions of each wire multiplied by their radial distance squared from the centre of mass position,

\begin{equation}
    \centering
    k = \sum_{i=1}^{3} \frac{m_i g}{L_i} R_{i}^2,
\end{equation}

where the static tensions in each wire, their radial distances from the centre of mass and their lengths are given as $m_i g$,  $R_i$ and  $L_i$, respectively. Unlike a single wire torsion balance, which uses the wire's elastic properties to provide the torsional stiffness, the torsional stiffness of the three-wire suspension is largely determined by its gravitational potential energy. The wire parameters were chosen such that the gravitational potential energy was much greater than the elastic energy stored in the wire due to deformation produced by simple pendulum oscillations \cite{SpeakeAnelasticity}. The Q factor should then not be limited by material losses in the wires.  The torsional resonance of the device was $f_0 = 0.88\,\rm{Hz}$ with a $Q \sim 220$ when in free swing (no active control) at pressure of 2\,mbar. The Q value is a factor of at least 10 less than expected and this maybe due to plastic deformations of the wires at the clamps. All measurements except those of the Q were taken at atmospheric pressure.

An optical lever which probed a mirror mounted at the geometric centre of the mass was used to measure the torsional motion of the mass relative to the suspension frame. The optical lever comprised a HeNe fibre-coupled laser (which was external to the vacuum system), a fibre-fed collimator, a lens and a quadrant photodiode (QPD). The collimator, lens and QPD were all fixed to the frame of the suspension. As mentioned above, tilts of the suspension frame result in a horizontal displacement of the mass relative to the frame. To reduce this coupling to the measurement system, the lens (a plano-convex lens of focal length $l_f = 100 \pm 2 \,\mathrm{mm}$), was nominally positioned at this distance from the QPD, resulting in rejection of translation motion from the readout. A schematic of the readout scheme is illustrated in Fig.\,\ref{fig:schematic}. The position of the lens relative to the QPD was adjusted to minimise the output for a given rotation of the mirror. Despite great care being taken in this process, inaccuracies in this distance could not be completely eliminated. A differential measurement was used to remove any residual coupling, described in Sec.\,\ref{sec:procedure}. A calibration of the QPD showed that, over the measurement range, the  QPD behaves linearly with a maximum non-linearity of 3\%.

As the system was adapted from the air-bearing discussed in Ref.\,\cite{GettingsSpeake}, the geometry was not suitable to implement the ILIAD interferometric rotation sensor \cite{ILIAD}.

A differential control scheme was implemented using 4 coil-magnet actuators, operating in pairs to damp the torsional mode of the mass. Permanent magnets were mounted on the lower plate of the torsion balance. However measurements of Q without the magnets ruled out eddy current damping as a dominant loss mechanism for the torsional mode. The differential control was used to damp the rotational mode and the rotational deflection was read out at DC by the QPD.

\section{\label{sec:procedure}Experimental Procedure and Results}

\begin{table}[htpb!]
    \centering
    \caption{Centre of mass shifts in the x and y axes due to the placement of the additional mass. Note that Positions 3 and 8 are used for subtractions for baseline measurements.}
    \begin{tabular}{c | C{2cm} | C{2cm}}
    \hline
    \hline
        & \multicolumn{2}{c}{centre of mass shift [mm $\pm 0.02\,$mm]} \\
        Mass Position & x & y \\
        \hline
        1 & -1.49 & 0\\
        2 & -0.73 & 0 \\
        3 & 0.12 & 0\\
        4 & 0.77 & 0\\
        5 & 1.52 & 0\\
        6 & 0 & -1.50\\
        7 & 0 & -0.73\\
        8 & 0 & -0.02\\
        9 & 0 & 0.68\\
        10 & 0 & 1.50\\
        \hline
        \hline
    \end{tabular}
    \label{tab:CoMPositions}
\end{table}

\begin{figure*}[t]
    \centering
    \begin{subfigure}{0.49\linewidth}
        \centering
        \includegraphics[width=\linewidth]{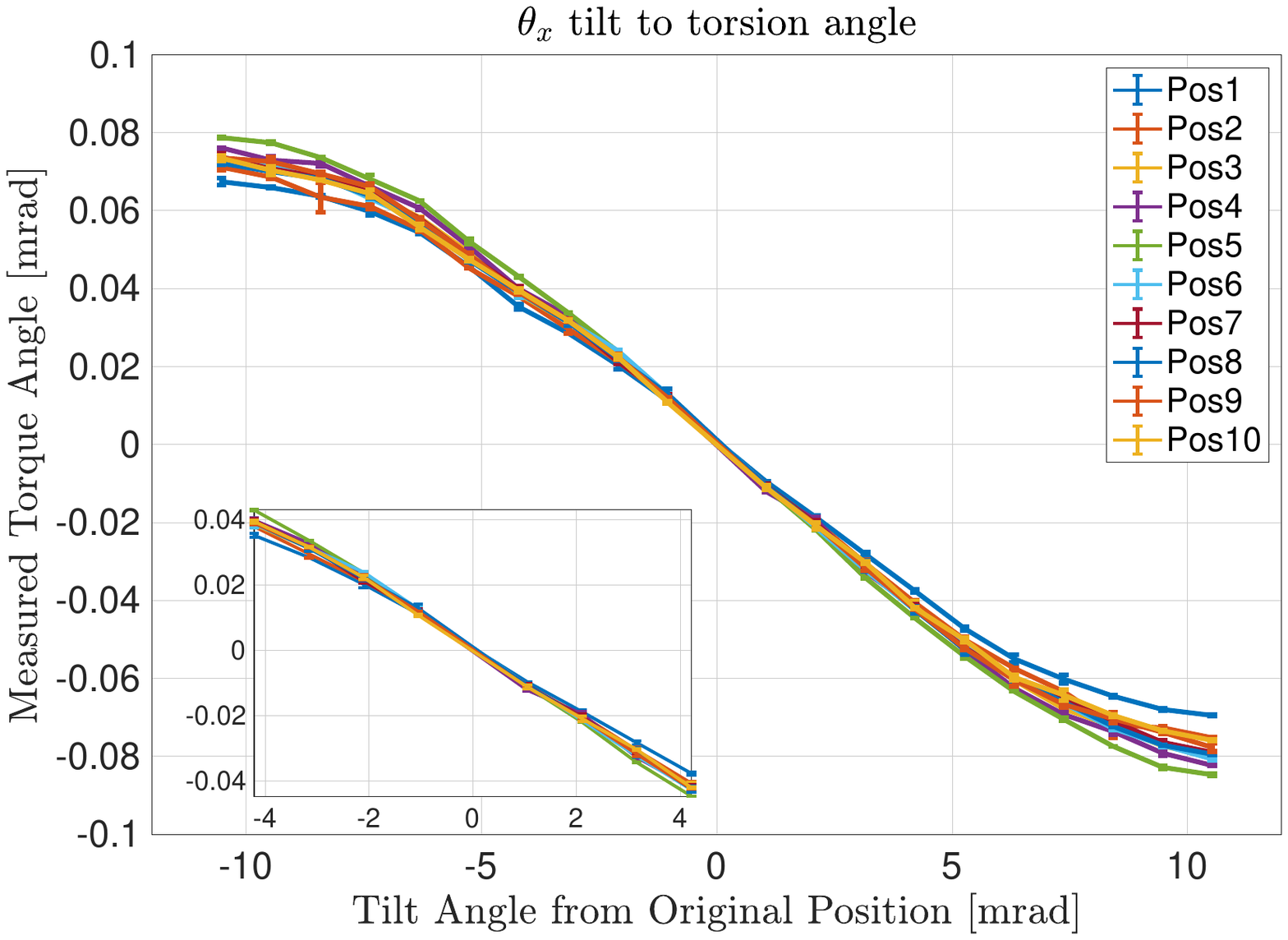}
        \caption{Tilt about x axis, $\theta_x$.}
        \label{fig:RawRX}
    \end{subfigure}
    \begin{subfigure}{0.49\linewidth}
        \centering
        \includegraphics[width=\linewidth]{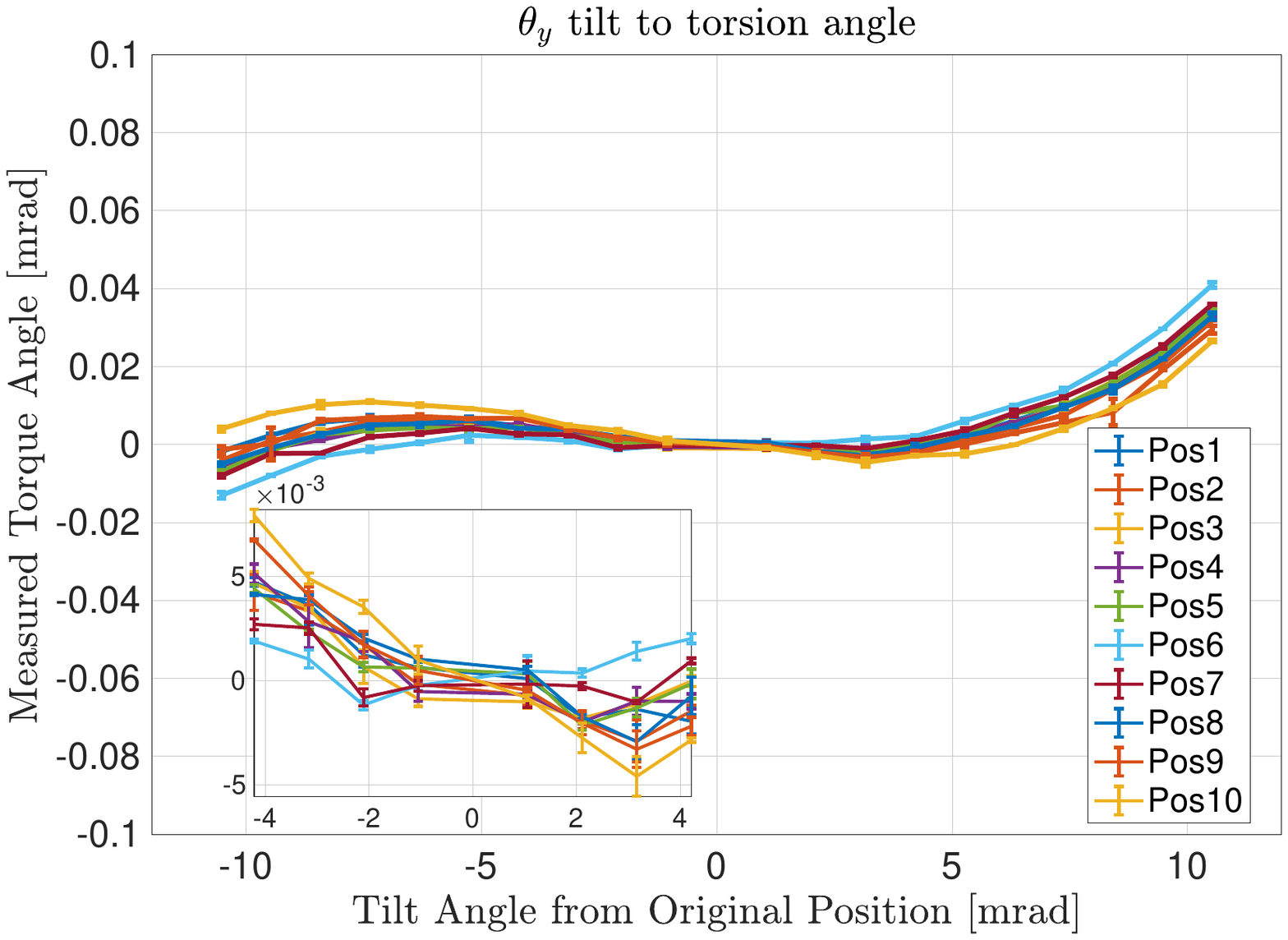}
        \caption{Tilt about y axis, $\theta_y$.}
        \label{fig:RawRY}
    \end{subfigure}
    \caption{Measured torsion angle for different mass positions due to DC tilt of the frame. The inset graphs show zoomed sections of the curve about the origin where the system is not dominated by the non-linearities at larger tilt angles. From this we determined that the differences in wire length from a common value were, $dL_1 = 0.17\,$mm, $dL_2 = 0\,$mm, and $dL_3 = 1\,$mm. This assumes that the lens is optimally located at its focal length.}
    
    \label{fig:RawTilt}
\end{figure*}

\begin{figure*}
    \centering
    \begin{subfigure}[b]{0.49\linewidth}
        \centering
        \includegraphics[width=\linewidth]{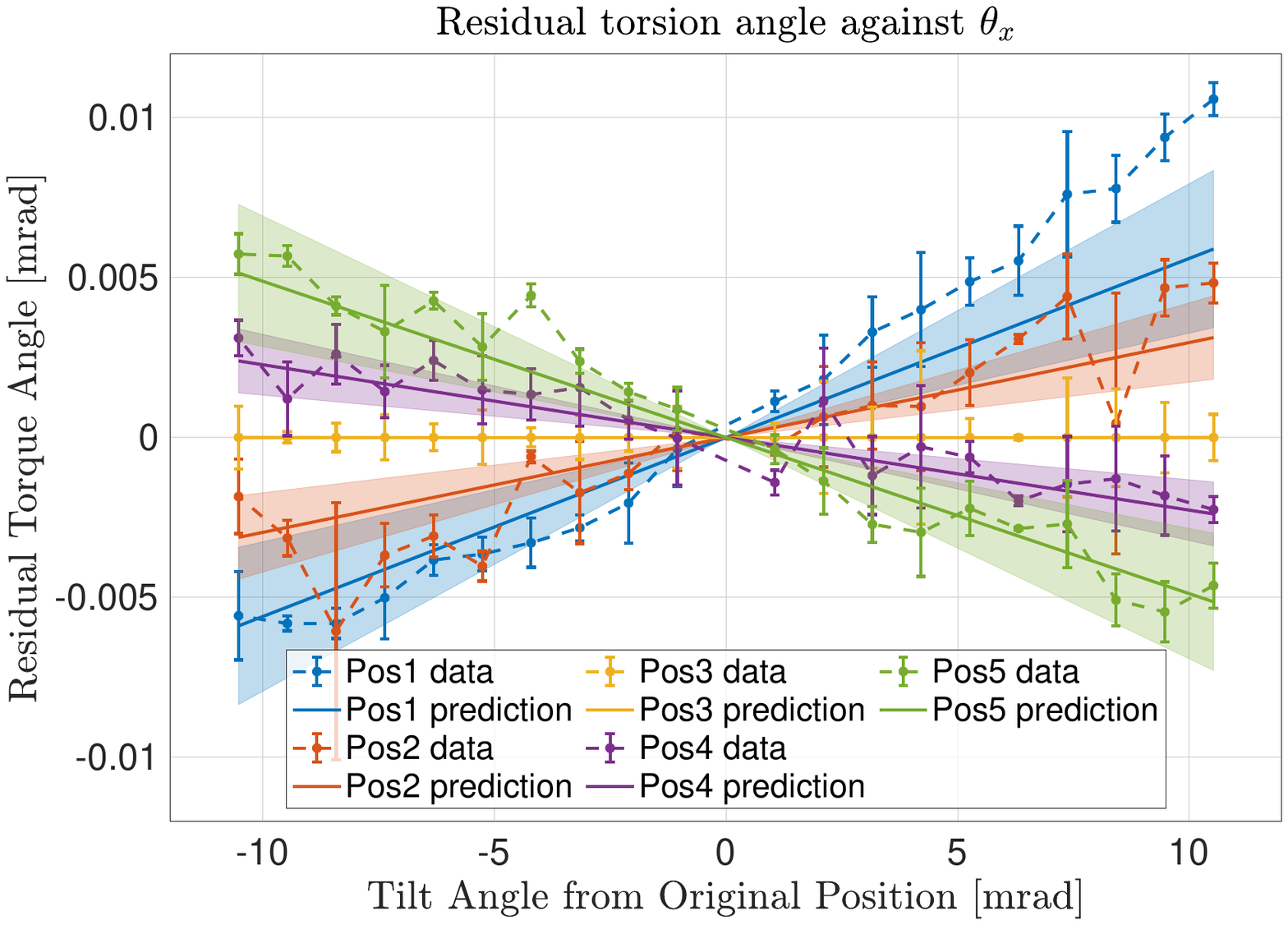}
        \caption{Tilt about x axis, $\theta_x$, for mass positions 1-5.}
        \label{fig:Xaff}
    \end{subfigure}
    \hfill
    \begin{subfigure}[b]{0.49\linewidth}  
        \centering 
        \includegraphics[width=\linewidth]{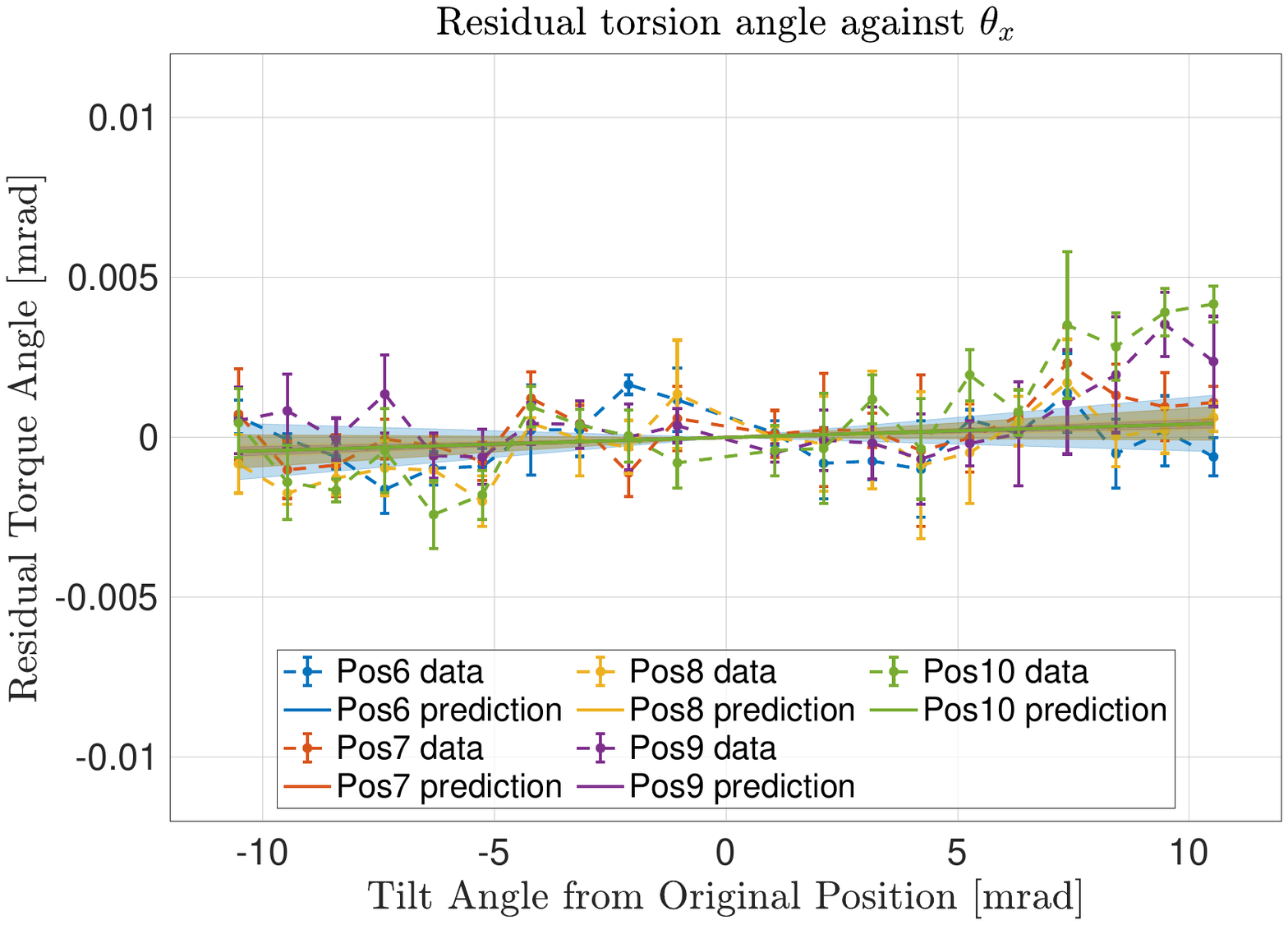}
        \caption{Tilt about x axis, $\theta_x$, for mass positions 6-10.}
        \label{fig:Xnon}
    \end{subfigure}
    \vskip\baselineskip
    \begin{subfigure}[b]{0.49\linewidth}   
        \centering 
        \includegraphics[width=\linewidth]{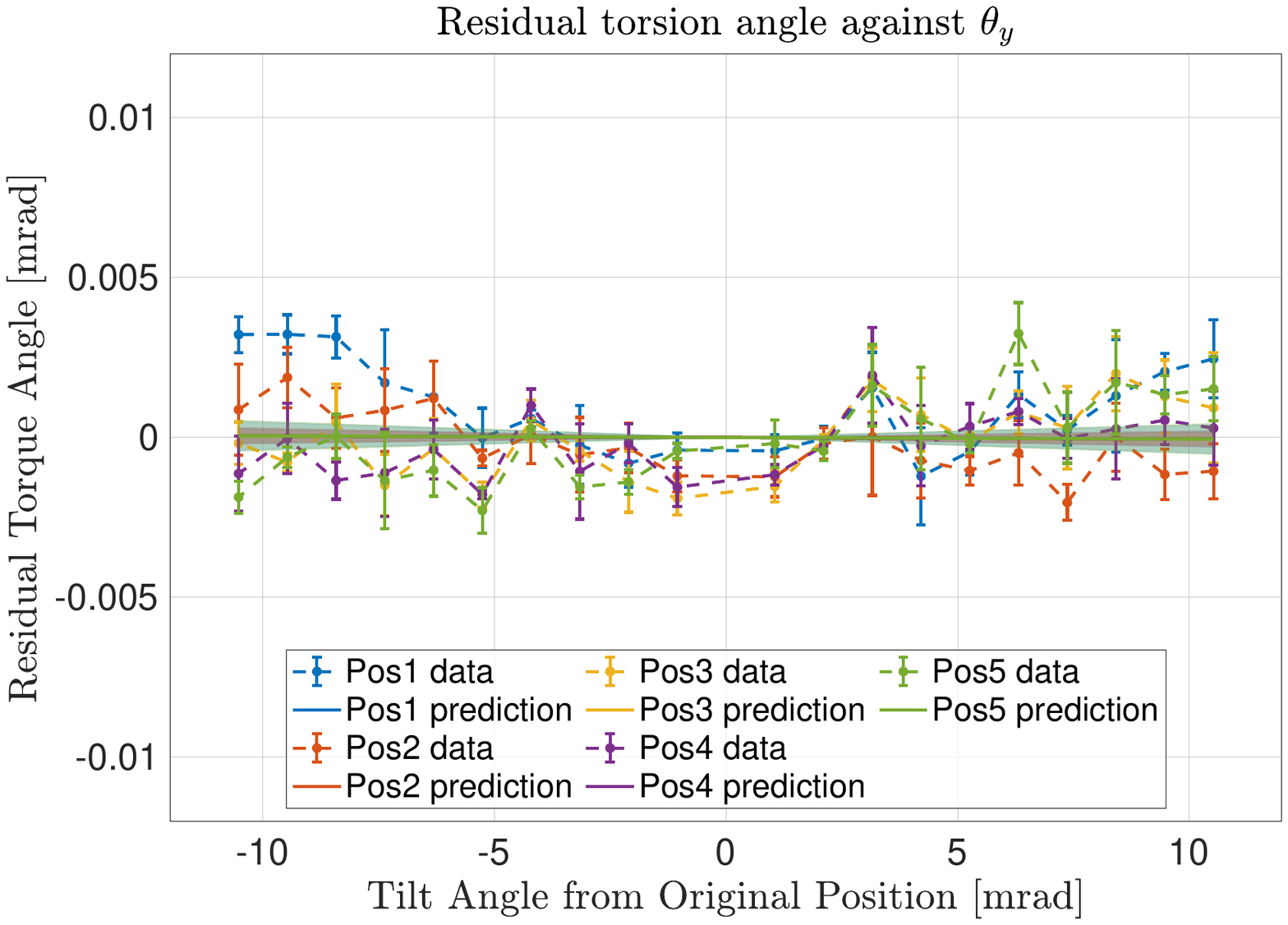}
        \caption{Tilt about y axis, $\theta_y$, for mass positions 1-5.}
        \label{fig:Ynon}
    \end{subfigure}
    \hfill
    \begin{subfigure}[b]{0.49\linewidth}   
        \centering 
        \includegraphics[width=\linewidth]{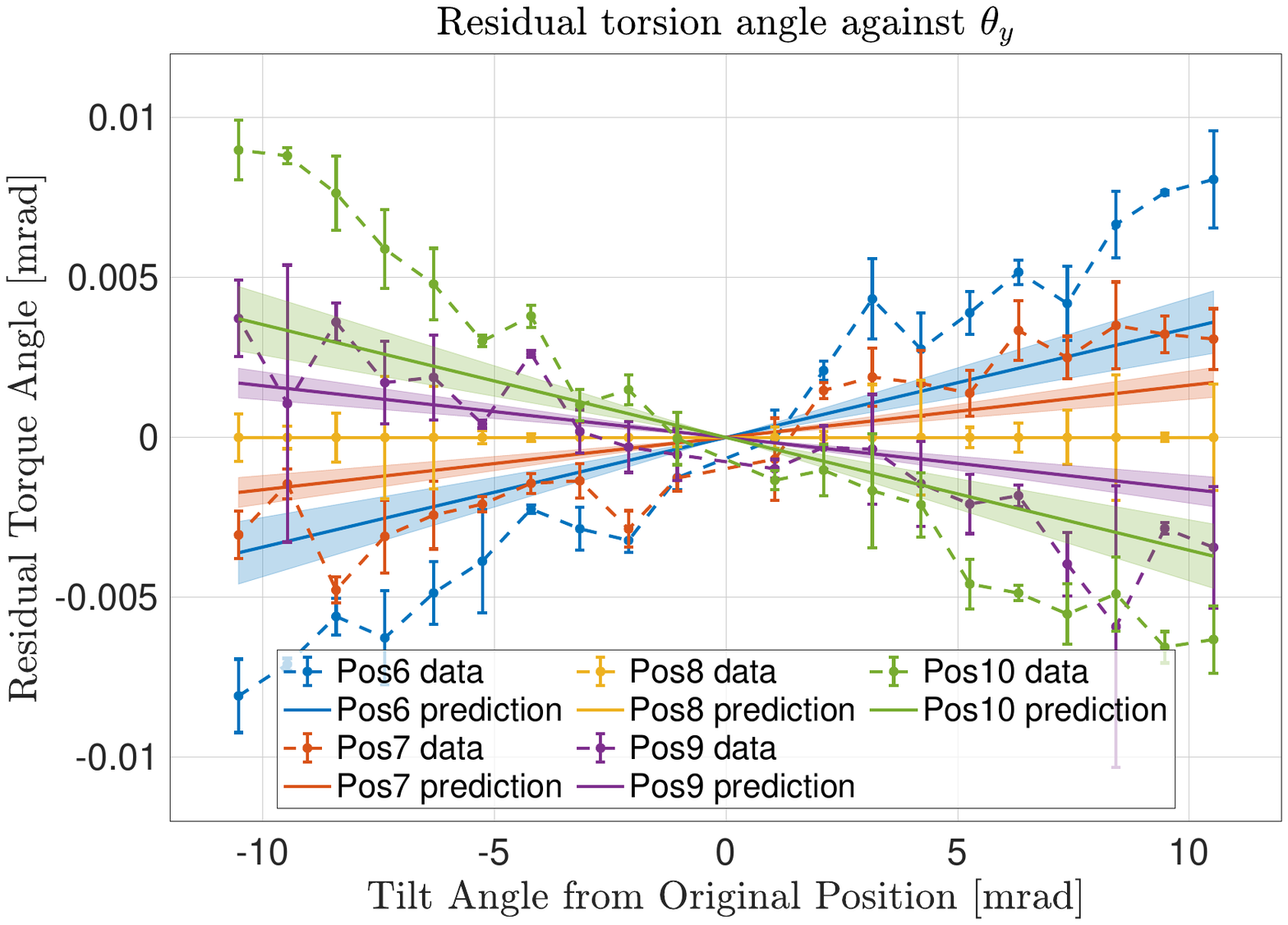}
        \caption{Tilt about y axis, $\theta_y$, for mass positions 6-10.}
        \label{fig:Yaff}
    \end{subfigure}
    \caption{Residual torsion angle due to placement of an additional mass. The central position (Position 3 or Position 8), is used as a reference. The dashed line shows the measured data, and the solid lines are the predictions made from our model. The filled section are the error on the prediction.}
    \label{fig:Subtractions}
\end{figure*}

The frame of the torsion balance was fitted with Thorlabs Z812 DC servo motor actuators which allowed remote control of  the frame's orientation in tilt about the X and Y axes. This enabled us to achieve micrometer level repeatability and accuracy in the positioning of the frame. A set of additional masses were located on the torsion balance to adjust the centre of mass position. This was equivalent to moving a 10g mass along the x and y axes of the coordinate system independently for each set of measurements. The mass positions are indicated in Fig.\,\ref{fig:schematic}. The effective centre of mass shift is shown in Tab.\,\ref{tab:CoMPositions}

A series of tilts were made moving the system back and forth between its original position and a new tilted position. This was done bidirectionally and for varying amplitudes of tilt. A baseline measurement was acquired for a specific mass position (see the caption in Tab.\,\ref{tab:CoMPositions}) which was then used as a reference for subsequent measurements. This was done independently for data taken for tilts about the x and y axes. Positions 1-5 (6-8) corresponded to tilts about the x (y) axes with position 3 (8) as the baseline measurement. This enabled us to eliminate the possibility of non-linearities being produced by the angular readout.

The measured torsion angle before subtraction is shown in Fig.\,\ref{fig:RawTilt}. Extensions of the wires can be determined from the tension changes due to the varying positions of the masses. The wires used for the suspension were tested and their vertical stiffness found to be $k_z = 2000 \pm 100 \,\mathrm{N/m}$ which was $\sim$ two-thirds of the theoretical value.
We use the this data  to calculate initial differences in the lengths of the wire prior to the loads being applied. Using the results of Sec.\,\ref{sec:PhysicalPrinciples} it can be shown that the residual coupling of tilt to rotation angle produced by differences in wire lengths is given, to first order, by 

\begin{equation}
    \phi = \frac{dL_1 + dL_2 - 2dL_3}{5 R}\theta_x + \frac{-2dL_1 + 2dL_2}{5 \sqrt{3} R}\theta_y,
    \label{eq:torsionangle}
\end{equation}

where $dL_i$ is the change in wire length compared to the baseline measurement, and $R$ is the radial distance to the wires from their geometric centre. We thus determined that the sensitivity to tilt of the torsion balance can be explained if the differences in the wire lengths from a common length are, $dL_1 = 0.17\,$mm, $dL_2 = 0\,$mm, and $dL_3 = 1\,$mm, assuming that the lens is optimally located at its focal length from the QPD. However, these length differences represent one possible solution to Eqn.\,\ref{eq:torsionangle}, where an arbitrary $dL_i$ can be chosen to be zero. Subtraction of the signals from the baseline measurements are shown in Fig.\,\ref{fig:Subtractions}.

We note that the non-linearities in both axes are similar in nature, occurring in the same direction and with similar amplitude. However, as the signal for tilts around the y-axis is much smaller, the non-linearities appear to have a larger affect on the measurement. This reinforces the fact that the non-linearities are not produced by the QPD. It is likely that the subtraction is less robust resulting in the residual torsion angle being larger than the linear model predictions given by Eqn.\,\ref{eq:torsionangle}. The error bounds in Fig.\,\ref{fig:Subtractions} are due to the uncertainty in the vertical stiffness, $k_z$, of the wires.

\section{Discussion}\label{sec:discussion}

\begin{figure}[htpb!]
    \centering
    \includegraphics[width = \columnwidth]{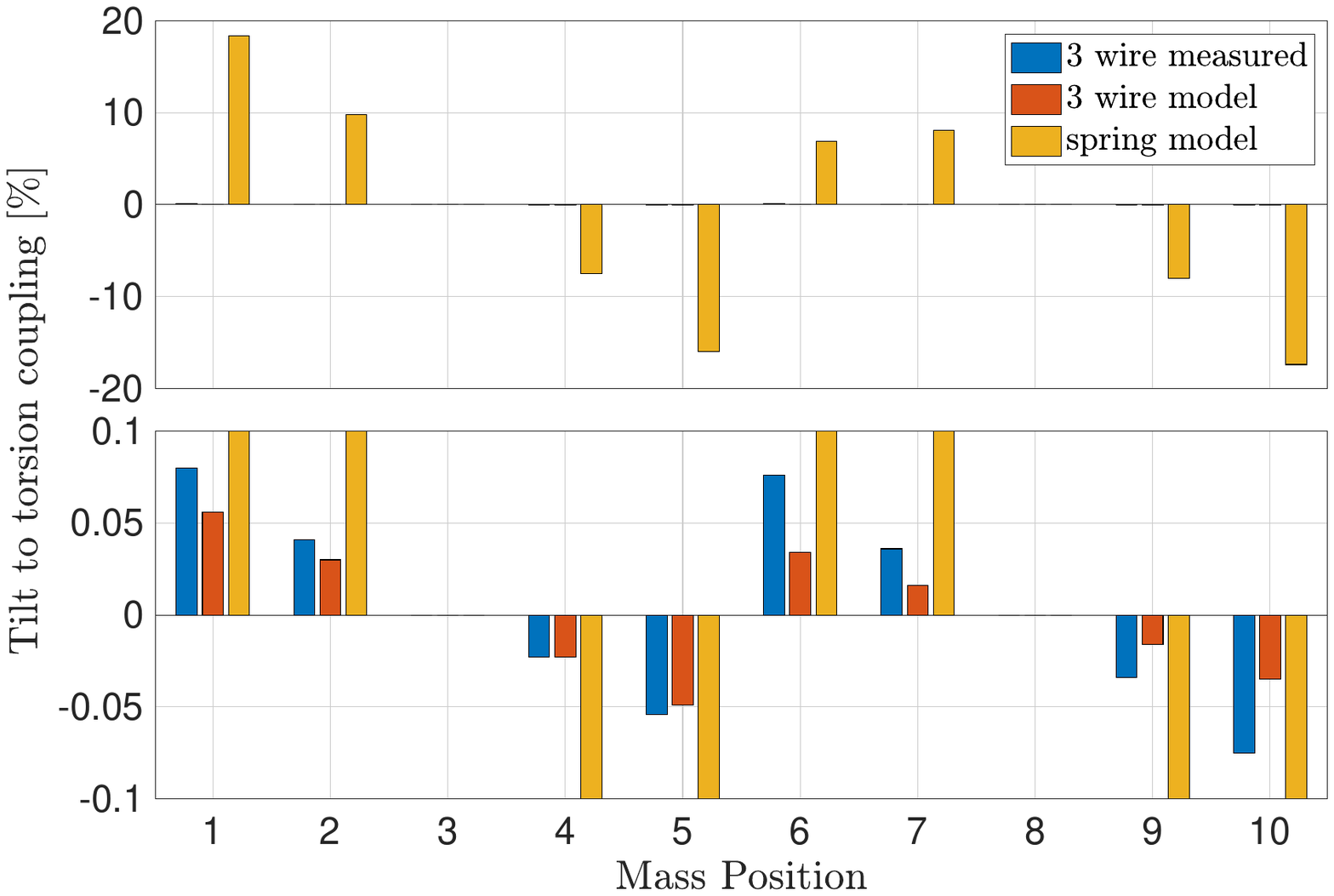}
    \caption{Tilt to torsion coupling comparison  due to the centre of mass offset from the additional mass. The blue bars represent the measured fit from Fig.\,\ref{fig:Subtractions} (a) and (d), the red bars are the predictions from the model, and the yellow curves are for the modelled case if there were springs of constant stiffness (independent of the position of the centre of mass). The upper tile shows the complete data, and the lower tile is zoomed in to show the smaller coupling due to the three-wire stiffness compensation.}
    \label{fig:CouplingComparison}
\end{figure}

It is useful to compare the results for the rotation produced by tilt given in Fig.\,\ref{fig:Subtractions} for the real torsion balance by comparing it with a device where the horizontal translational stiffness is not given by the tensions of the support wires, that necessarily are determined by the position of the centre of mass of the lower plate, but by springs of fixed stiffness. This system could be realised using the air-bearing apparatus described Ref.\,\cite{GettingsSpeake}. If we assume that these stiffnesses are given as $k_1 = k$, $k_2 = k$ and $k_3 = 2k$, and we vary the mass positions as shown in Fig.\,\ref{fig:schematic}, the residual torsion angles are much larger. A comparison is shown in Fig.\,\ref{fig:CouplingComparison}. We achieve {at least a factor of 200 suppression of the tilt-to-torsional coupling using the three-wire method where the stiffness compensates for the change in the centre of mass position. 

The data shown in Fig.\,\ref{fig:RawTilt} shows a non-linear response of the torsion balance to tilts. This behaviour cannot be described by imperfections in the optical lever system which is sufficiently linear as mentioned in Sec.\,\ref{sec:experimental_design}. We have considered the clamping mechanism for the wires as a potential source of these non-linearities. Tilts and accelerations of the system would cause the wires to partially wrap around the rollers, changing the effective length of the wires. However this would be common for all tilt measurements and would be removed via subtraction of the baseline measurement. Also the magnitude of these contributions are much too small to fit the observations. There are two likely candidates for these non-linearities. Plastic deformations of the wires could occur during clamping which changes the elastic characteristics of the wires, and result in non-linear changes in the wire lengths as the tensions vary due to centre of mass changes and tilts. We would also expect that changes in the deviations from cylindrical symmetry in the three wires at the upper clamps would produce tilt sensitivity and this mechanism could be non-linear with tilt angle. 


Although a torsion balance measures angular rotations about the vertical axis, the angle is dependent on the torsional stiffness of the device. It is appropriate to consider the limitation to the measuring capability introduced by tilts. We will consider a comparison between a traditional torsion balance used for measurements of the inverse square law of gravity outlined in Ref.\,\cite{HoyleThesis,HoylePaper}, and a three wire suspension. Note an extra layer of isolation is usually implemented in torsion balances, where a prehanger wire is used to reduce the coupling from external disturbances to the torsional measurement. As mentioned in Sec.\,\ref{sec:Introduction}, in the case of Ref.\,\cite{HoyleThesis,HoylePaper}, the 80\,cm long, $20\,\mathrm{\mu m}$ diameter torsion fibre was isolated via a 3\,cm long 150\,$\mathrm{\mu m}$ diameter prehanger fibre. The torsional stiffness for this device was $3.17 \times 10^{-9}\,\mathrm{Nm/rad}$ which is 7 orders of magnitude softer than the suspension described in this paper, with a tilt to torsion angle coupling of 1.6\% resulting in a tilt to torque noise of $5 \times 10^{-11}\,\mathrm{Nm/rad}$. For our prototype system we find, for the raw tilt to torsional coupling, the noise is $3 \times 10^{-4}\,\mathrm{Nm/rad}$. However the sensitivity of our device is limited due to the wide separation of the wires and the difficulties in ensuring equal static wire lengths. A dedicated device for use in short-range force style experiments such as the Casimir force could achieve torsional stiffnesses of $3 \times 10^{-3}\,\mathrm{Nm/rad}$, and with wire length differences of $\pm 10\,\mathrm{\mu m}$ this would result in the tilt to torque angle coupling of 0.01\%, corresponding to a torque noise of $3 \times 10^{-7}\mathrm{Nm/rad}$. Thus the tilt to torque noise and the inherent sensitivity of the traditional torsion balance is superior to that of a three wire torsion balance. 

\begin{table*}[htpb!]
    \centering
    \caption{A comparison of Casimir experiments from the 2003 AFM style experiments by R.S. Decca et al \cite{Decca2003}, Lamoreaux's 1996 torsion balance \cite{Lamoreaux} experiment, and a proposed three-wire torsion suspension.}
    \begin{ruledtabular}
    \begin{tabular}{c|c|c|c|c|c}
         Device type & Interaction area, $A$ $[\mathrm{m^2}]$ & Nominal separations, $a$ [$\mathrm{\mu m}]$ & Normal Force, $F$ $[\mathrm{pN}]$ & Torque, $\tau$ $[\mathrm{fNm}]$ & Stiffness, $k$ $[\mathrm{Nm/rad}]$ \\
         \hline
         AFM & $\sim 3 \times 10^{-10}$ & 0.5 & $\sim 6$ & $\sim 1$ & $8.6 \times 10^{-10}$ \\
         Torsion balance & $\sim 5 \times 10^{-7}$ & 2 & $\sim 40$ & $\sim 2000$ & $2.4 \times 10^{-6}$ \\
         Three-wire & $\sim 5 \times 10^{-4}$ & 10 & $\sim 65$ & $\sim 10000$ & $3 \times 10^{-3}$ \\
    \end{tabular}
    \end{ruledtabular}
    \label{tab:comparisons}
\end{table*}

However, most experiments used to measure Casimir-like forces use planar-spherical geometries, such as the torsion balance utilised by Lamoreaux \cite{Lamoreaux}, and atomic force microscope (AFM) like devices such as those utilised by Decca et al. \cite{Decca2003}. 

The three wire torsion balance compensates its sensitivity via the orders of magnitude greater interaction areas involved using planar-planar geometry, and the use of patterned masses to generate a transverse Casimir force \cite{Chalkley2011}. This transverse force generates a torque on the three wire torsion balance. A comparison of these devices is shown in Tab.\,\ref{tab:comparisons}. 

We also note that, if required, the rotational stiffness can be reduced using electrostatic fields between suitable geometries of electrodes.

\section{Conclusions}\label{sec:conclusions}

We present a model and experimental investigations for determining the coupling sensitivity of input tilts and horizontal accelerations to the torsional mode of a three-wire torsion balance. The final purpose of such a device is to enable the parallelisation of two planar objects for measurements of forces at micrometer and sub-micrometer separations without the requirements of active control in the tilt degrees of freedom.

The investigation was aimed at measuring the tilt to torsional coupling due to the position of the centre of mass. The prototype suspension discussed demonstrates a coupling of <1\% in the raw tilt to torsional signals and non-linearities for large tilt angles. These large tilt angles are 3 orders of magnitude greater than the expected RMS tilt of the ground which is approx $\sim 1\,\mathrm{\mu rad}$. Assuming a 1\% coupling this would result in an RMS noise of $10\,\mathrm{nrad}$ for the large wire length offset in this suspension. Reducing the differences in wire length would suppress this coupling even further, with careful design the length differences between the wires could be $<10\,\mathrm{\mu m}$.

To remove the effect of the static wire length differences and any common systematic noise, we used baseline measurements and subtracted the remaining datasets from them. This enabled us to view the effects only due to the centre of mass position and the wire length differences due to the changes in tensions. The residual tilt coupling to the torsional mode after subtraction was <0.1\% and removed some of the non-linearities that appear in the measurement. The purposeful centre of mass offset results in differences in the wire lengths for the suspension due to the changes in tension as described by Eqn.\,\ref{eq:torsionangle}. The residual difference in the wire lengths are $<50\,\mathrm{\mu m}$. To achieve the measured torsion angle before subtractions shown in Fig.\,\ref{fig:RawTilt}, the initial difference in the wire lengths are constrained to be $\leq 1$\,mm, however their actual values are difficult to determine due to the non-linearities and residual couplings in the measurement scheme. The asymmetry in the system was amplified by the offset of the geometric centre from the wires from the centre of mass of the suspended plate. We found that this has an  minimal effect and only changes the numerical coefficients in Eqn.\,\ref{eq:torsionangle}, further highlighting the stiffness compensation of the three-wire suspension. The first order coupling of the wire length differences is the dominant coupling factor of tilt or horizontal accelerations to the torsional mode of the suspension.

Further improvements could be made to the set up which may potentially reduce the non-linearities. First, the use of a dedicated angular sensor such as an autocollimator or a custom sensor such as an angular interferometer \cite{ILIAD} which should have a higher rejection of translational motion. Secondly, a robust method of clamping the wires to their nominal lengths and tensions before suspending the reference mass would ensure the wire length differences are minimised. This would reduce the likelihood of plastic deformations during the suspension procedure and thus the direct coupling of tilt to twist in individual wires.

After suspension of the device, because the linear tilt sensitivity potentially depends on both the wire length differences and their deviations from axial symmetry at the attachment points, the net tilt sensitivity could be eliminated via changes in the centre of mass position. This is a result of the changes in tension, and thus the changes in the lengths of the wires.

We have recently completed the construction of a 3-wire torsion balance with associated mechanisms for achieving close proximity of planar test and source surfaces and an interferometric rotation detector. In a future publication we will report on the performance of this instrument.

Given the simplicity of the mechanical systems that we have studied here, it is possible that our analysis may find other applications, for example in gravitational wave suspensions \cite{Robertson_2002, Aston_2012}.

\section{\label{sec:acknowledge}Acknowledgements}

We would like to thank Arthur O'Leary and Hugo Yamagushi for their initial work, as undergraduates, on the three-wire suspension. We are also grateful to John Bryant for his invaluable assistance and members of the mechanical workshop at University of Birmingham.
One of us (CCS) would like to acknowledge the late Professor James E. Faller for his enduring inspiration.

We are grateful for support from EPSRC UKRI through an award of a New Horizons grant in 2022. 

\bibliography{main.bib}

\end{document}